\pdfoutput=1
\documentclass[aps,prl,twocolumn,groupedaddressm,showkeys]{revtex4-1}
\usepackage{graphicx} 
\usepackage{epstopdf} 

\begin{document}


\title{Coupling Human Mobility and Social Ties}


\author{Jameson L. Toole}
\email[]{jltoole@mit.edu}
\affiliation{Engineering Systems Division, MIT, Cambridge, MA, 02144}

\author{Carlos Herrera-Yaq\"ue}
\affiliation{Department of Applied Math, Universidad Polit\'ecnica, Madrid, 28040, Spain}

\author{ Christian M. Schneider}
\affiliation{Department of Civil and Environmental Engineering, MIT, Cambridge, MA, 02144}

\author{Marta C. Gonz\'alez}
\affiliation{Department of Civil and Environmental Engineering, MIT, Cambridge, MA, 02144}

\begin{abstract}
Studies using massive, passively data collected from communication technologies have revealed many ubiquitous aspects of social networks, helping us understand and model social media, information diffusion, and organizational dynamics.  More recently, these data have come tagged with geographic information, enabling studies of human mobility patterns and the science of cities. We combine these two pursuits and uncover reproducible mobility patterns amongst social contacts. First, we introduce measures of mobility similarity and predictability and measure them for populations of users in three large urban areas. We find individuals' visitations patterns are far more similar to and predictable by social contacts than strangers and that these measures are positively correlated with tie strength.  Unsupervised clustering of hourly variations in mobility similarity identifies three categories of social ties and suggests geography is an important feature to contextualize social relationships.  We find that the composition of a user's ego network in terms of the type of contacts they keep is correlated with mobility behavior. Finally, we extend a popular mobility model to include movement choices based on social contacts and compare it's ability to reproduce empirical measurements with two additional models of mobility.
\end{abstract}

\maketitle

The rise of ubiquitous mobile computing has facilitated the generation, collection, and storage of massive data sets of human behavior. Social interactions are captured in calls, emails, and tweets, while movement is logged by check-ins and GPS traces~\cite{Onnela2007, Gonzalez2008, Schneider2013, Toole2012}. Studied separately, social and mobility data have produced a wealth of insights. Our understanding of information and diseases spread~\cite{Ugander2012, Vespignani2011, Watts2005}, how our friends affect our well being~\cite{Centola2011,Christakis2007}, and how societies are structured~\cite{Eagle2010,Watts2002, Ratti2006, Eagle2009, Bettencourt2013, Batty2008} has been greatly improved by studying large social networks. Mobility data has revealed that human movement is regular, predictable~\cite{Song2010science, Song2010natphys}, and unique~\cite{DeMontjoye2013}. To complement empirical findings, a number of simple models have been proposed to reproduce the basic dynamics of both social networks~\cite{Newman2003, Boguna2003, Watts1998} and mobility~\cite{Song2010natphys, Grabowicz2013, Simini2012, Schneider2013}, but the two have been traditionally treated as independent.

Recognizing the interaction between social behavior and mobility, researchers began measuring correlations between the two. They found that social networks are heavily influenced by geography.  We are far more likely to be friends with someone nearby than far away~\cite{Liben-Nowell2005}, a fact that is useful for predicting missing links~\cite{Wang2011, Crandall2010}.  With an estimated 15\% to 30\% of all trips taken for social purposes, it is not surprising that the movement of our friends can improve predictions of where we will be next~\cite{Cho2011, Domenico2012, Grabowicz2013}.  While insightful, the primary interest of most previous studies was measuring and reproducing patterns of geographic distance and it's impact on network topologies~\cite{Grabowicz2013}. In dense urban areas, however, distance is less restrictive. Residents have access to a variety of transportation options and are free to choose locations that provide the best goods and services rather than the closest.  The self-organized districts and neighborhoods of cities make it more natural to describe mobility as movement between sets of locations, or habitats~\cite{Bagrow2012}. Which habitats users share with their contacts and when they share them may indicate the nature of the social relationship: e.g. a coworker or a friend~\cite{Eagle2009PNAS}.  Two individuals co-located between 9am and 5pm on weekdays likely have a different relationship than two who are found in the same area at midnight on a Saturday. In these scenarios, mobility is defined and measured as discrete visits to places within a city that are shared with different types of social contacts at different times and previous work has shown that users who visit similar places are more likely to be friends in online location based social networks~\cite{Cho2011}.


Here we describe a set of metrics to explicitly measure patterns of mobility and social behavior that occur within the context of cities. Using call detail records (CDRs) produced by millions of mobile phone users, we find that individuals have far more similar visitation patterns to social contacts than to strangers and that the movement of these contacts can be used to reconstruct a considerable portion of the individuals' movements.  We also find strong correlations between tie strength and mobility similarity and show that mobility similarity can be used to classify social relationships and recover semantic information about the nature of a link in the social network. Finally, we propose an extension to the mobility model described in \cite{Song2010natphys} that incorporates movement based on the visitation patterns of social contacts and can reproduce empirical relationships found in the data. We call this model the GeoSim model and compare it against empirical data and two other mobility models. The generality of these results is demonstrated by their reproducibility in three different cities in two different countries. This study presents advances in the understanding of how social behavior affects our spatial choices in the context of information and communication technologies (ICTs).

\begin{figure}
 \includegraphics[width=1\linewidth]{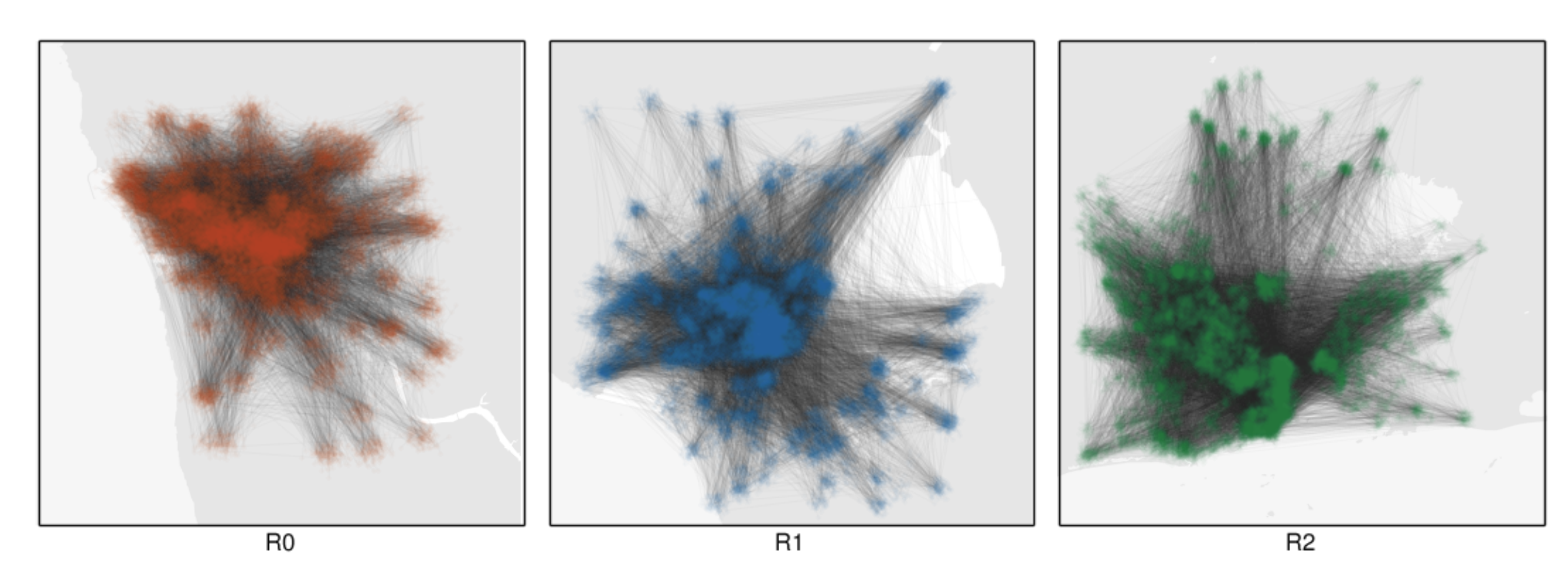}%
 \caption{\label{fig:activity_map}A small sample of calls between residents is shown for each of three cities.  CDRs provide the location of each caller as well as record of communication between then.  A dot is drawn at the approximate location of a user and a link appears between two users calling each other. Our aim is to identify useful and reproducible patterns from this coupled tangle of social and spatial behavior.}
 \end{figure}
 
\section{Materials and Methods}

 \begin{figure*}[t!]
 \includegraphics[width=1\linewidth]{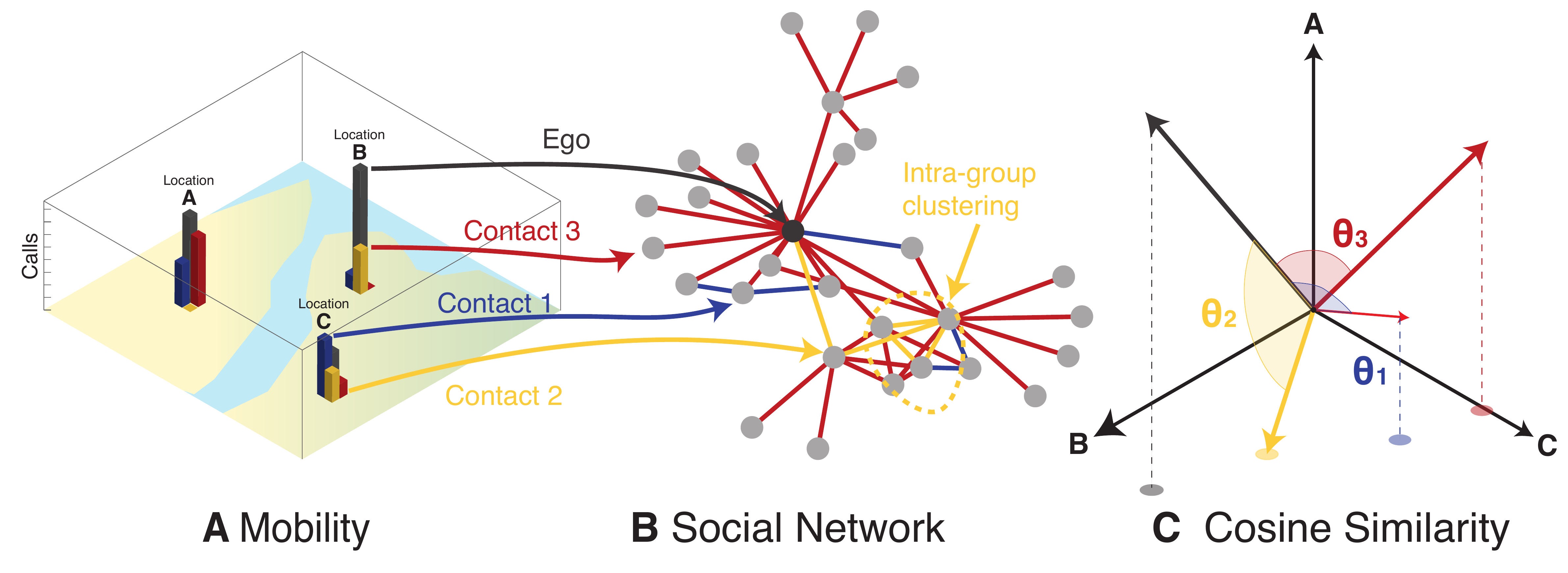}%
 \caption{\label{fig:similarity_diagram} Similarity of visitation patterns between nodes in social networks. For each user, we keep track of (A) how many visits are made to locations across the city and (B) construct a social network by tracking calls to others.  We can then define (C) the geographic cosine similarity between two users by computing the cosine of the angle between any two vectors in the location space.}
 \end{figure*}

\subsection{Data}
Call detail records (CDRs) are generated when a mobile phone user performs an action that requires the provider's network, for example placing a call or sending a text message.  These records generally contain the ID of the tower the phone connected through, which gives a rough estimate of the user's location.  When the individual receiving a call or message is a customer of the same provider, the unique identifier of the receiver and their location may also be stored. CDRs allow us to observe mobility patterns of individuals and construct social networks containing millions of people.  Figure \ref{fig:activity_map} shows a small sample of calls between city residents during a single hour and illustrates dynamics of the urban system we wish to understand.

Our data consist of anonymized CDRs collected from three cities (R1, R2, and R3) in two different industrialized countries.  Two cities (R1 and R2) were obtained from the same provider in country 1, while another provider was used for the third city (R3).  The observation period covers 15 months in R1 and R2 and 5 months in R3 and contains over 1 billion events in total.  Each record provides the time of the communication event, an anonymous unique ID for the caller and callee, and the ID of the tower used by at least the caller (in the case of R3) and in some cases the callee (R1 and R2).  More information on the datasets can be found in the electronic supplementary materials (ESM).

\subsection{Social and Mobility Measurements}
In each city, we construct a social network containing all users (nodes) with sufficient call volume and connect users (edges) if they have regular contact between each other (see ESM for more detail).  Each node is assigned a $48\times L$ location matrix $\mathbf{L}$, where $L$ is the number of unique cell towers in the city.  Each row of this matrix corresponds to an hour of a typical weekday and hour of a typical weekend day (giving 48 hours in total) and each element $L_{t,j}$ contains the number of times that a user made a call from location $j$ during hour $t$ across the entire observation period (Figure \ref{fig:similarity_diagram}A).  We refer to individual rows of this matrix $\mathbf{v(t)}$ as \textit{location vectors}. The location matrix and location vectors can be used to compute various mobility properties of nodes (mobile phone users). Summing all elements of the location matrix gives the number of calls made and received by a user $N= \sum_{t,j}L_{t,j}$ while summing each column and dividing by $N$ provides the frequency of visits a user made to every location in the city, $f_j = \frac{1}{N}\sum_t L_{t,j}$.  Summing visits to each location at all times gives a single location vector $\mathbf{v}$ for each user and represents the total visits made to each location over the period of data collection. Applying the sign function and summing across all elements of this vector provides the number of unique locations visited $S = \sum_j sign(v_j)$.  All of these features are measures of a user's mobility behavior within the city.

We can also compare the location matrices and vectors of two mobile phone users and measure similarities between the two.  While a number of metrics could be used to measure mobility similarity between nodes (Figure \ref{fig:similarity_diagram}B), here we focus on the cosine similarity between the location vectors of two nodes $i$ and $j$ defined as: $\cos\theta_{i,j} = \frac{\mathbf{v_i}\cdot \mathbf{v_j}}{|\mathbf{v_i}||\mathbf{v_j}|}$.  The cosine similarity measures the cosine of the angle between two vectors in our $L$-dimensional \textit{location space} (Figure \ref{fig:similarity_diagram}C).  It has been shown to correlate strongly with the probability of being friends in an online social network~\cite{Cho2011} and has a number of desirable properties. It is sensitive to visit frequencies rather than set intersections alone, so two users who share frequently visited locations appear more similar than those who share less important destinations. Unlike the Pearson correlation coefficient, it does not overstate similarity when vectors contain many zero elements (as is often the case) and finally, the cosine similarity is a measure of the angle only and is not affected by differences in the total number of calls made by two users.  For the remainder of this paper, we refer to the cosine similarity between two locations vectors as {\it mobility similarity}.

The mobility similarity between two users can be computed from their entire movement history or visits during a small portion of a weekday or weekend.  In the former case, we assign a single mobility similarity value to an edge in the network, while in the latter, we assign a time series of cosine similarity $\cos \theta(t) = \frac{\mathbf{v_i(t)}\cdot \mathbf{v_j(t)}}{|\mathbf{v_i(t)}||\mathbf{v_j(t)}|}$.  This time series reveals how often two users visit the same places at a given time of the day and will later function as an attribute to differentiate between types of social contacts.

Within this mathematical framework, we can calculate an upper bound on how much of an individual's location vector can be reconstructed from a linear combination of the location vectors of other users.  For example, a co-worker may share office space with an individual, but not live in the same neighborhood, while the opposite may be true for a member of that individual's family.  By combining the visitation patterns of the co-worker and family members, however, a complete picture of an individual's visitation patterns can be obtained.  Mathematically, we define a set of users $F$ for each individual $i$ in the network.  For example, we may choose $F$ to be neighbors in $i$'s ego network or a random set of nodes.  The location vectors $\mathbf{v_j}$ where $j \in F$ are used as columns of an $|F|\times L$ matrix we denote as $\mathbf{A}$ and span a subspace of the $L$-dimensional location space.  We then use QR-decomposition to find an orthonormal basis $B = { q_1,\dots, q_{|F|}}$ for $\mathbf{A}$.  Our target user's location vector is then projected into this vector subspace: $\hat{\mathbf{v}} = \sum_{i=1}^{|F|} \langle q_i, \mathbf{v} \rangle q_i$. This projection represents the best approximation of a user's visits based on the visits of users in $F$.  We can quantify how it compares to a user's true visitation patterns by taking the ratio of it's magnitude with the magnitude of the actual location vector $|\mathbf{v}|$.  We refer to this ratio as {\it predictability} and define it mathematically as $\frac{|\hat{\mathbf{v}}|}{|\mathbf{v}|}$.  When predictability is 1, the visitation frequencies of a user can be completely obtained from location vectors of users in $F$ and when it is 0, nothing about their visits can be learned. We note that for values between 0 and 1, predictability cannot be interpreted as the fraction of a user's visits that can be recovered as the vector norms are computed using the standard L2 norm. In principal, however, these two quantities should be strongly correlated because the individual elements location vectors can never be negative.

We next apply these methods and metrics to social network and mobility data from three cities.

\section{Results}
\subsection{Correlations between social behavior and mobility}
\begin{figure*}[t!]
\includegraphics[width=1\linewidth]{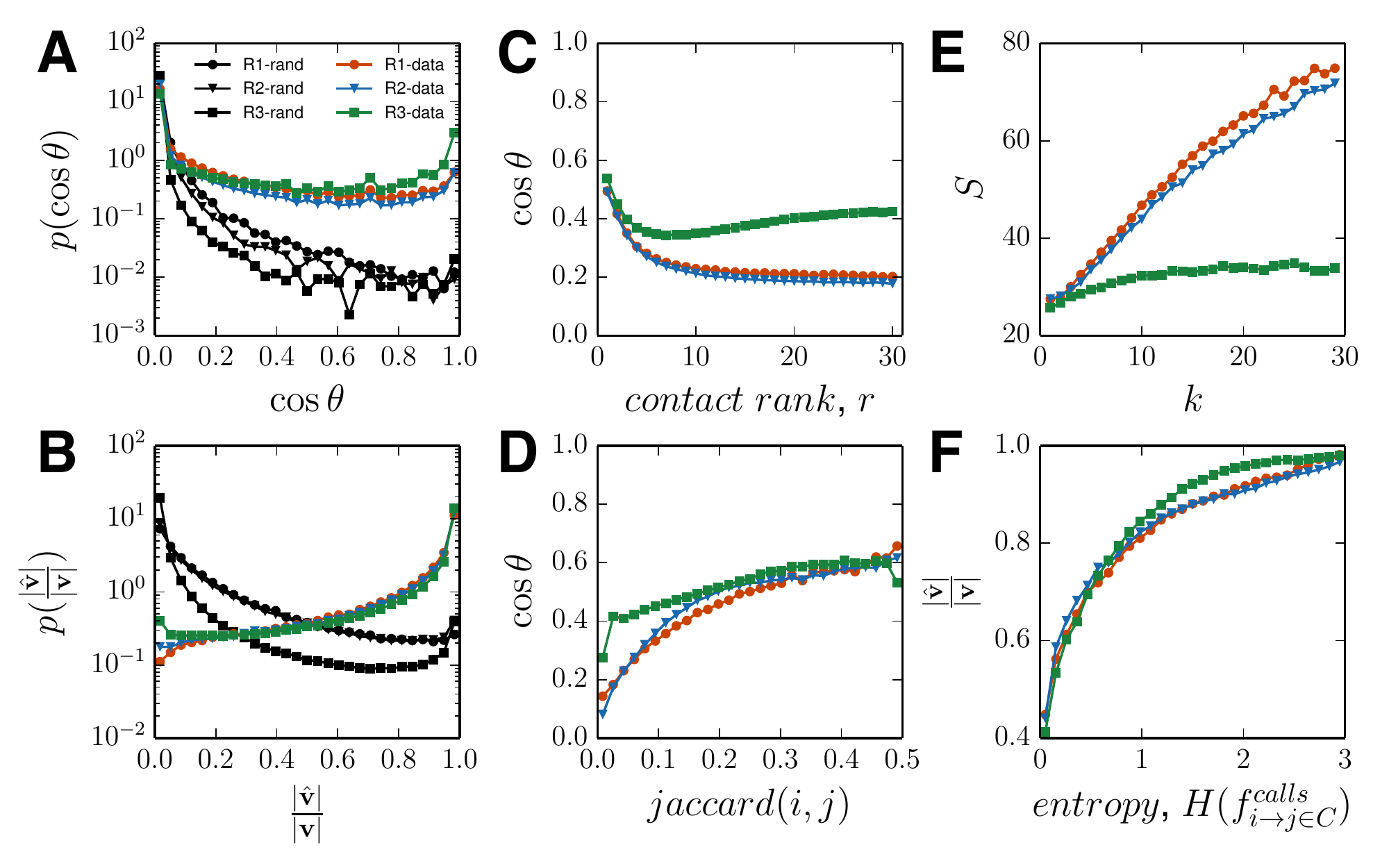}%
\caption{\label{fig:correlations}Correlations between mobility and social behavior. For each city, we compute  the (A) distribution of cosine similarity and (B) predictability using observed edges (colored lines) and compare to distributions made using randomized edges.  We find both mobility similarity and predictability are much higher when using actual social contacts compared to random users.  Social similarity is also correlated with mobility similarity. (C) Ranking each user's contacts by number of calls, we find that stronger ties are more geographically similar.  (D) Moreover, the more common contacts shared by two users, the more geographically similar those individuals tend to be.  Finally, we explore how social behavior is correlated with mobility.  (E) We find that users with more unique contacts tend to visit more unique locations.  (F) Users who distribute their calls to contacts more evenly (higher entropy) are more predictable than users with more uneven call distributions.  This suggests that users who share social attention more evenly also share locations.  Figure S2 and S3 in the ESM show these results controlling for call frequency.}
\end{figure*}

Though similarity can be measured between any two arbitrary nodes and predictability from an arbitrary set of nodes $F$, we hypothesize that an individual will likely be more similar to and predictable by social contacts.  To test this, we compare the mobility similarity between users that call each other regularly with the similarity between random users and the predictability achieved using a node's social ties with the predictability using random sets of nodes (essentially rewiring the social network, but leaving mobility intact).  Figures \ref{fig:correlations}A and \ref{fig:correlations}B show the distribution of similarity and predictability values for the networks in each city.  We find significantly more similarity and predictability in empirical networks when compared to random re-wirings.  The similarity distribution is bimodal, with peaks at very low similarity near 0 and very high similarity near 1. We measure very high values of predictability when using an individual's social contacts as opposed to a random set of people in the same city.  As other studies have suggested, we find that visitation patterns are strongly linked to our social relationships; our movements are far more similar to our social contacts than random users.

Interestingly, we observe higher levels of mobility similarity between users separated by short network distances.  We find that two connected nodes are on average 10 times more geographically similar that two randomly selected nodes. Nodes separated by two hopes, or ``friends of friends", are nearly twice as similar as randomly selected nodes and this elevated similarity is observed up to three hops from an individual (see ESM Figure S5 for details).  This result is expected as two users who do not contact each other may both visit the same friend.  
 
Next, we explore the relationship between tie strength and mobility similarity.  We rank all contacts in each user's ego network by the number of calls shared between them (1 being contact that shares the most calls) and compute the average mobility similarity for all edges with a given rank (Figure \ref{fig:correlations}C).  Stronger contacts have higher mobility similarity on average than weaker ties, though this effect subsides for contacts below rank 10. We note that region R3 shows a slightly different trend. This is likely due to the shorter observation period in this region resulting in few individuals with more than 10 regular contacts, biasing the tail of this distribution (see ESM for more details).  We also observe a positive correlation between social similarity as measured by the Jaccard index between the neighbors of two nodes and mobility similarity (Figure \ref{fig:correlations}D); individuals who share more social contacts share more locations.  

We also find other aspects of social behavior to be correlated with mobility. Individuals with more friends tend to visit more locations, but despite this exploratory behavior, are still more predictable due to increased information provided by additional contacts to reconstruct these movements from (Figure \ref{fig:correlations}E). Again R3 appears as an outlier due to the shorter observation period and the absence of mobility information on the user receiving a call. We then measure the entropy of the distribution of frequencies that a user $i$ calls another contact $j$ and find that individuals with more entropic  calling patterns (distribute their calls more evenly) also visit more unique places and are more predictable (Figure \ref{fig:correlations}F).  The visitations patterns of those who spread social attention more evenly can be more easily reproduced.  Finally, to ensure that these results are not an artifact of sampling frequencies, we compute these distributions and correlations controlling for the number of CDR events by and the degree of a user, finding no change in the relationships (see Figures S1, S2, and S3 in the ESM).

\subsection{Contextualizing social contacts with mobility}
\begin{figure}[!t]
\includegraphics[width=0.5\linewidth]{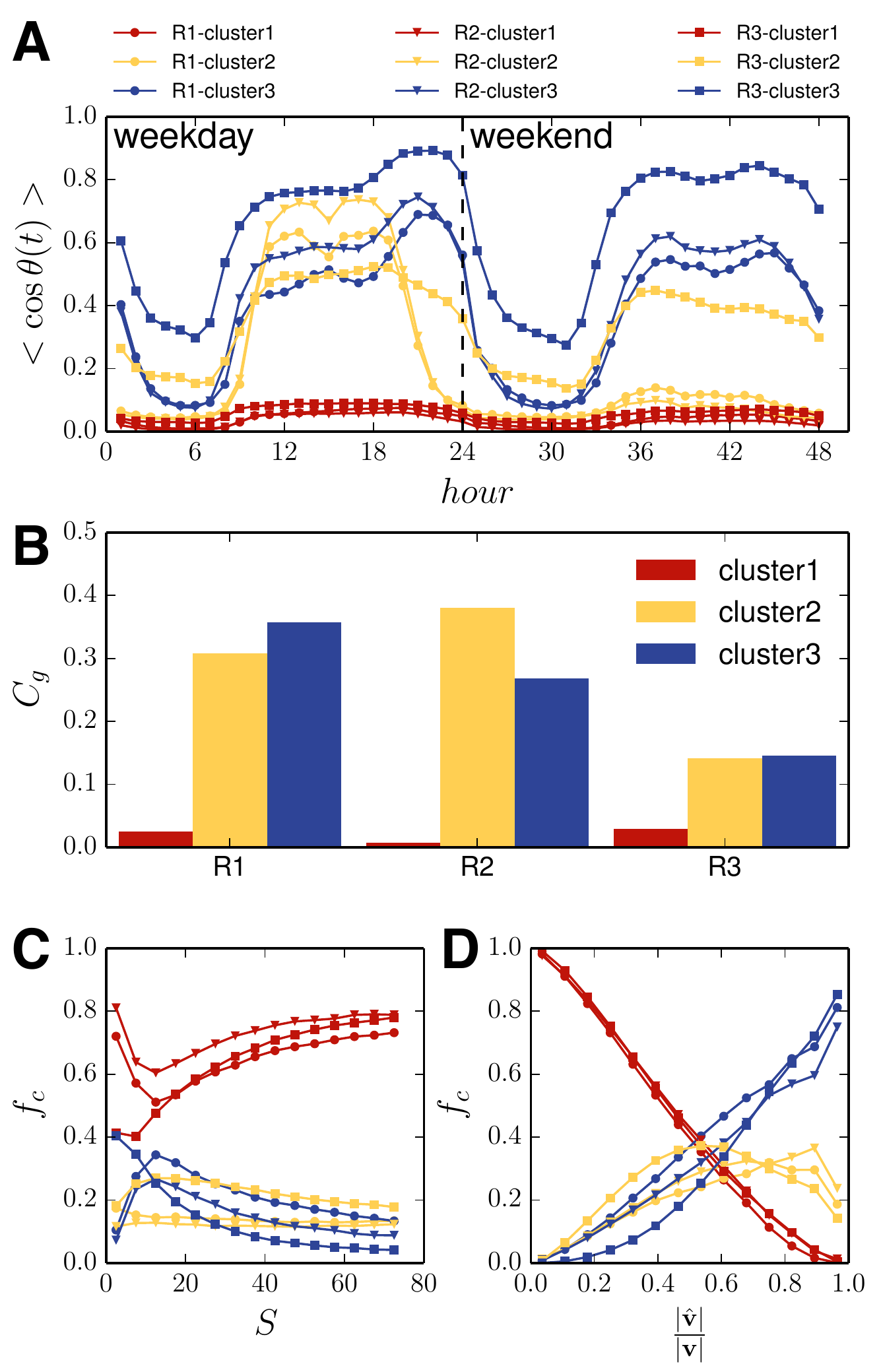}%
 \caption{\label{fig:clustering} Characterizing social ties based on similarity of movement over time. (A) We perform k-means clustering on the set of similarity time series from edges in the network.  We find three groups emerge in each city:  ($i$) {\it acquaintances} who have low levels of similarity across all times, ($ii$) {\it co-workers} who have elevated similarity during work hours on weekdays, but lower levels on weekends, and ($iii$) {\it family/friends} who have high similarity on nights and weekends. (B) For each city we construct subgraphs containing only edges in a single cluster.  We find that these subgraphs retain high clustering coefficient ($C_{g}$) within the co-worker and family/friend group while acquaintances are far less likely to have ties among each other. Finally, we explore how an user's behavior correlates with the mobility characteristics of their immediate social network. (C-D) We group nodes based on their mobility characteristics (unique locations visited $S$ and predictability $\frac{|\hat{\mathbf{v}}|}{|\mathbf{v}|}$) then compute the fraction of edges that belong to each of the identified clusters for each node in the group.  Individuals that are more exploratory (visit more unique places) tend to have higher fractions of {\it acquaintances} ties than individuals with lower mobility while the reverse trend is observed for the most predictable individuals.}
 \end{figure}

Having demonstrated that social behavior and location choices are strongly correlated, we next use temporal variations in mobility similarity to provide context into the type of social relationship between two individuals in our networks.  We measure mobility similarity $\cos \theta(t)$ over the course of a typical weekday and weekend under the hypothesis that different types of social contacts will have different levels of similarity at different times.  To identify any groups, we use a simple k-means unsupervised clustering algorithm on these similarity time series.  We find three persistent groups.  While we have no ground truth data about the nature of these relationships, for clarity, we label each group according to it's qualitative signature: ($i$) {\it acquaintances} with uniformly low levels of similarity, ($ii$) {\it co-workers} with high similarity during work hours on weekdays and low similarity on nights and weekends, and ($iii$) {\it family/friends } with high similarity on nights and weekends.  Figure \ref{fig:clustering}A shows the cluster centers for each group.  While other interesting clusters are found for $k>3$, they appear as subgroups of the three general archetypes we discuss here.  More information on the clustering method along with results for different numbers of clusters and different clustering methods can be found in the ESM.  These three groups appear in each city despite the unsupervised nature of the algorithm; cluster centers start at random locations, yet find remarkably similar final positions in each city.

Assigning each edge to a cluster based on the time series of mobility similarity effectively paints all edges in the next in a specific color as illustrated above in Figure \ref{fig:similarity_diagram}B.  Previous work has found that edges in real social networks are much more likely to be arranged in triangles, resulting in high clustering coefficients.  In this case, we expect that some social groups, such as co-workers or close friends, should exhibit high degrees of intra-group clustering, while others such as acquaintances do not. For example, many of an individual's {\it co-workers} visit similar places during work hours and tend to call each other because they are part of the same office community. We find evidence of this when measuring the clustering coefficient within subgraphs containing only edges belonging to a single mobility similarity cluster (Figure \ref{fig:clustering}B).  Interestingly, the clustering coefficient ($C_{g}$) of {\it acquaintances} is much lower than the {\it co-workers} and {\it family} ties despite consisting of nearly 70\% of links in the network.  This provides additional evidence that we are capturing very different types of relationships with our classifications based on mobility similarity. Moreover, these results highlight mobility similarity as a property to label functional communities within social networks as well as individual edges. 

Next, we consider how the composition of an individual's ego-network correlates with their mobility.  Is a person with a stable job and family is likely to be less exploratory and more predictable than a young college student with many acquaintances? To answer this, we bin nodes into groups based on two mobility metrics, the number of unique locations visited  $S$ and how predictable that user is $\frac{|\hat{\mathbf{v}}|}{|\mathbf{v}|}$.  We then compute the fraction of edges that belong to each classification for all nodes in each mobility bin.  Figure \ref{fig:clustering}C shows that users who tend to visit more unique locations tend to have a higher fractions of {\it acquaintances} in their ego network, while Figure \ref{fig:clustering}D suggests that less predictable individuals tend to have fewer contacts in this category. Conversely, less spatially explorative individuals and individuals that are easier to predict tend to have higher fraction of {\it co-workers} and {\it family/friends} labels in their ego network. These results again show the ability of mobility similarity to add contextual attributes to a network and reveal novel relationships between the structure of a user's ego network and their mobility behavior.  In future works, it may be interesting to explore correlations between the mix of one's ego network and social behaviors such as their propensity to form new contacts \cite{Miritello2013}.

\subsection{Coupling social ties and mobility}
Given the clear empirical relationship between social contacts and mobility, our remaining task is to identify a coupled model that captures these dynamics. While a number of models consider mobility alone~\cite{Song2010natphys, Simini2012, Gonzalez2008}, only a few have attempted to link the two~\cite{Grabowicz2013, Cho2011}. Those that have combined social and mobility behaviors have consistently found  nearly 15-30\% of trips are made for social purposes. Though these coupled modeled have had considerable success reproducing patterns of geographic distance within social network structure, but, as we show, do not always capture properties of geographic similarity.

In light of the time scales we are studying, we make the assumption that our social network is static and extend the mobility model introduced by Song et al. \cite{Song2010natphys} to include movement choices based on social contacts.  We call our extension the \textit{GeoSim} model\footnote{We have released code and data required to run this model online at http://humnetlab.mit.edu/wordpress/downloads.}. We compare our model to the original individual-mobility model (IM model) by Song et al. and the Travel-Friendship model (TF model) described by Grabowicz et al.  See ESM for more details on implementation and parameters for model comparisons.

The GeoSim model works as follows: first, a population of $N$ agents are initialized and connected to replicate the undirected social network constructed from the CDR data in R1.  Each edge that exists in the call data, exists in the model, but all weights and similarities are set to 0.  Agents are randomly assigned to a location at the start and their location vectors are initialized to reflect this single visit.  They are allowed to move in a discrete space of $L$ locations replicating the towers from CDRs.  

Each time step corresponds to a single hour of the day. At each time step, individuals decide whether or not to change locations according the waiting time distribution measured in \cite{Song2010natphys}, a power-law with an exponential cutoff $p(\Delta t)=\Delta t^{-1-\beta}\exp(\Delta t / \tau)$ where $\beta=0.8$ and $\tau=17$ hours.  If an individual moves, they must decide to either return to a previously visited location with probability $1-\rho S^\gamma$ or explore and visit a new one with probability $\rho S^\gamma$, where $S$ is the number of unique locations they have visited thus far and $\rho=0.6$ and $\gamma=0.6$ are parameters chosen by procedures outlined in\cite{Song2010natphys}.  In the original model, an individual $u$ {\it preferentially returns} to a location $l$ with probability proportional to the frequency of previous visits, $P(l) \propto f_l^u$  and new locations to explore are chosen uniformly at random (note that in our version of the model distance is irrelevant). 

\begin{figure}[h]
\includegraphics[width=1\linewidth]{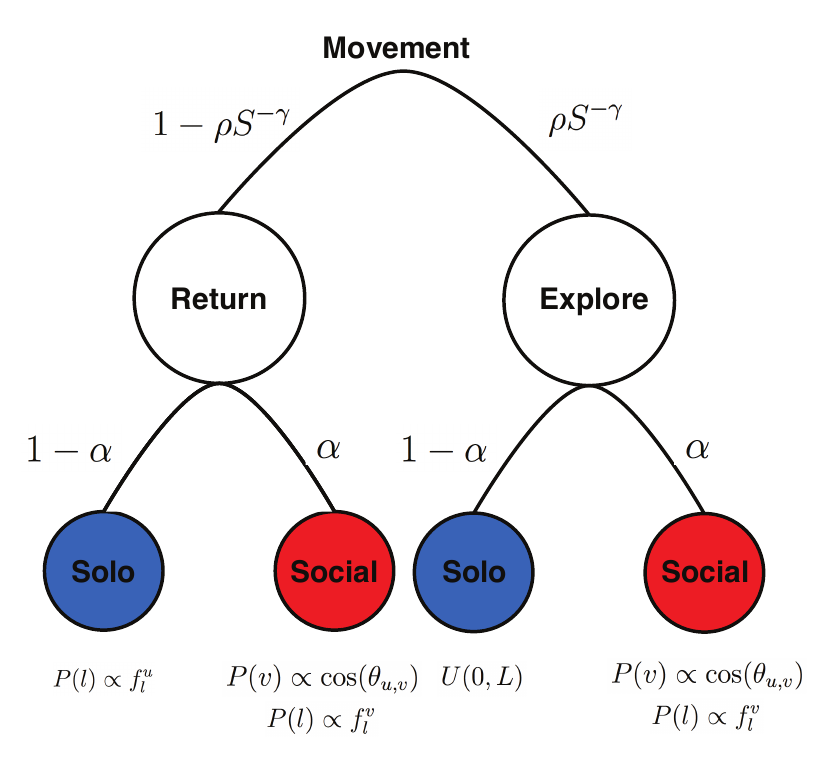}%
 \caption{\label{fig:model_description}A schematic description of the GeoSim model.  As in the IM model presented by Song et al., individuals first decide whether to return to a previously visited location or explore a new location.  The actual choice of location to visit, new or returning, is made based on either a social influence with probability $\alpha$ or individual preference with probability $1-\alpha$.}
 \end{figure}

In our extension of this model, we choose some locations based on social influence.  When picking a return location, our agent has two possibilities. With probability $1-\alpha$, they select a return location with the preference for locations they have visited in the past as in the original model. With probability $\alpha$ a social contact $v$ is chosen.  The probability a given contact is chosen is directly proportional to the current mobility similarity between the two, $P(v) \propto \cos(\theta_{u,v})$ and a location to visit is chosen based on a preference to visit locations frequented by the selected contact, $P(l) \propto f_l^v$ (note the location choice is repeated until an agent finds a location they have visited before).  In the social case, this amounts to preferential return based on a contact's visit frequency as opposed to the ego's visits.  In the event that an agent is exploring a new location, the same weighted social coin is flipped.  This time, though, with probability $1-\alpha$ a random, previously unvisited location is selected and with probability $\alpha$ the agent again chooses a contact based on mobility similarity and chooses a new place to visit based on the visit frequencies of that contact.  The cosine similarity across all edges is computed and updated over as the model progresses and changes dynamically during the simulation.  A schematic of this process can be found in Figure \ref{fig:model_description}.
  
In this variant of the mobility model, the parameter $\alpha$ controls the influence of social contacts on the visitation patterns of individuals.  When $\alpha = 0$, we recover the original mobility model of \cite{Song2010natphys}, while when $\alpha=1$ all location choices are influenced by social ties.  In reality, each user may have an inherent value of $\alpha$ that we cannot observe.  To incorporate this heterogeneity, we simulate this model for a number of distributions of the parameter $\alpha$. We find an exponentially distributed $\alpha$ with a mean of $\langle \alpha\rangle=0.2$ produces a close fit to distributions of mobility similarity and predictability observed in the population and refer the reader to the ESM for results for different distributions of $\alpha$. This value is consistent with the results of both Cho et al.~\cite{Cho2011} and Grabowicz et al.~\cite{Grabowicz2013} who find that roughly 15-30\% of trips were motivated by social intentions.

\begin{figure}[h]
\includegraphics[width=1\linewidth]{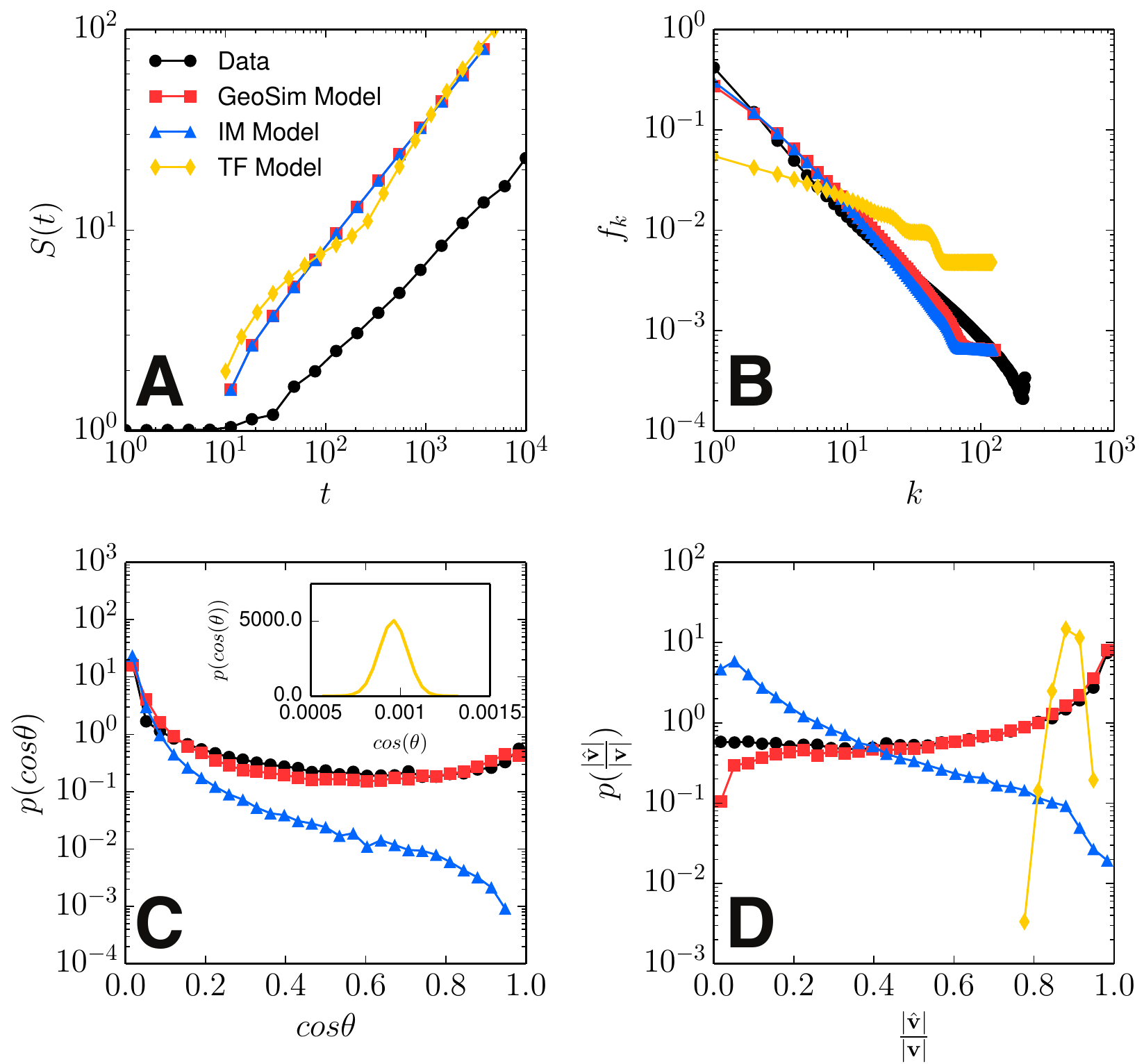}%
 \caption{\label{fig:model_compare}Comparing social mobility models.  A) We compare model results simulating the rate of exploration $S(t)$ compared to empirical data.  While all three models appear to estimate more absolute locations visited, the rate of this growth is consistent between them and in-line with data.  B) For each user, we sort locations based on the number visits and compute the frequency that a user visits a location of rank $k$.  We find that the IM models and our extension to it reproduce this distribution well, while the TF model is much flatter, distributing visits more evenly over all locations. C) Only the GeoSim model is able to reproduce patterns of mobility similarity and D) predictability.  The TF model results shown in the inset in C shows similarity values orders of magnitude below the observed data.  As the similarity is heavily influenced by the frequency distribution of visits, this deviation is likely due to the flatter distribution of $f_k$ produced by the TF model.}
 \end{figure}

Having found an appropriate distribution for $\alpha$, we next compare simulation results with this distribution to results from the IM model (equivalent to the GeoSim model with $\alpha=0$) and the TF model all run for the same 1 year duration and populations size.  Like the IM model it extends, the GeoSim model is able to reproduce elements of individual mobility such as the rate of exploration of new locations $S(t)$ over time (Figure \ref{fig:model_compare}A) as well as frequency at which users visit their locations $f_k$ (Figure \ref{fig:model_compare}B).  Here the TF model adequately reproduces exploration rates, but produces a flatter visit frequency distribution.  In the case of mobility similarity and predictability, however, only the GeoSim model reproduces observed behavior (Figure \ref{fig:model_compare}C-D).  Interestingly, the TF model results in relatively high predictability of users, despite similarity values orders of magnitude lower than those observed in the data or with the IM model.  This is likely due to the flattened frequency distribution which the cosine similarity is highly sensitive to.  Even if two users share a few locations due the friendship component of the TF model, there are preferential dynamics that will continually bring those two users back to that place, increasing cosine similarity.  On the other hand, this flat frequency distribution makes it highly likely that users will share at least some locations in commons with each other, making it possible to reproduce location vectors based on social contacts.  Despite it's inability to recover these distributions, the TF model is the only model tested that builds a social network endogenously.  For this reason, we hope future work will find variants on this model capable of dynamically reproducing empirical data of both social and mobility behavior.

\section{Discussion}
Linking mobility to social ties has generated a number of insights into the dynamics of both.  Social networks are embedded in geography where face-to-face interactions are often preferred and chance of interacting with those nearby is greatest.  At the same time, we are willing to travel to achieve this proximity and rendezvous at places across the city for work and play.  Novel high resolution data sets passively collected from mobile, online devices now enable us to quantify the correlation between mobility similarity and social behavior.  Here we have offered new metrics and empirical findings that relate social behaviors to mobility similarity and predictability.  Our results show that our mobility is far more similar to our social contacts than strangers and that this similarity can be used to reconstruct our own mobility patterns. We find strong, positive correlations between tie strength and mobility similarity. Moreover, temporal variations in this similarity reveal three distinct groups of social ties that hint at semantic types of relationships such as co-worker or family member.  These subgraphs often have high levels of intra-group clustering, suggesting functional groups of individuals within the network. The mix of these groups amongst the edges of an individual's ego network is correlated with their mobility behavior; users with many dissimilar contacts tend to explore more locations. Speaking to their generalizability, these results persist across three different cities in two countries.

Finally, we extended an established mobility model to include choices based on social behavior that replicates the empirical findings described here as well as from other works. We call this model the GeoSim model and have compared its results to two similar models.  We hope that this model provides a useful tool for future work in the area.  The findings presented have a number of implications for those interested in social networks or mobility applications extracted from ICTs. Additional contextual information of relationships may help predict missing links or provide critical details to more accurately model of the flows of information or diseases.  Urban planners or those needing good estimates of travel demand can incorporate social mechanisms like the ones described here to improve on their models and to capture movements previously unaccounted for.  Robust findings that classify social contacts from passive data alone may influence future studies and help with data informed policies through city science.  In the new data rich reality of cities, deeper insight into the connections between us will help make the places we live more sustainable, efficient, productive, and fun.

\section{Authors Contributions}
Jameson L. Toole designed and performed data analysis and wrote the manuscript.  Carlos Herrera-Yag\"ue designed and performed data analysis.  Christian M. Schneider designed data analysis.  Marta C. Gonz\'alez coordinated the study.  All authors gave final approval for publication. 

\section{Acknowledgments}
\begin{acknowledgments}
This work was partially funded by the BMW, the Accenture-MIT alliance and the Center for Complex Engineering Systems (CCES) at KACST under the co-direction of Anas Alfaris. Jameson L. Toole would like to acknowledge funding awarded by a National Science Foundation Graduate Research Fellowship.
\end{acknowledgments}

\bibliography{geo_social}

\end{document}